\documentclass[manuscript=article, layout=twocolumn]{achemso}
\setkeys{acs}{articletitle = true}

\usepackage[T1]{fontenc}
\usepackage[utf8]{inputenc}
\usepackage{graphicx}
\usepackage{amsmath}
\usepackage{color}
\usepackage{comment}
\usepackage{hyperref}
\usepackage{multirow}
\usepackage{subfig}
\usepackage{bm}
\usepackage{braket}
\usepackage{siunitx}
\usepackage[normalem]{ulem}

\DeclareSIUnit{\kcal}{\kilo\cal}

\title{Mitigating Errors in Variational Quantum Eigensolver Calculations through Extended Random Phase Approximation}

\author{Michael Wagner}
\altaffiliation{Department of Physics, Faculty of Nuclear Sciences and Physical Engineering, Czech Technical University in Prague, Břehová 7, 115 19 Prague 1, Czech Republic}
\affiliation{Heyrovsk\'{y} Institute, Czech Academy of Sciences, Dolej\v{s}kova 3, 18223 Prague 8, Czech Republic}

\author{Pavel Begy}
\altaffiliation{Department of Physical Chemistry, Faculty of Science, Palacký University Olomouc, 17. listopadu 1192/12, 779 00 Olomouc, Czech Republic}
\affiliation{Heyrovsk\'{y} Institute, Czech Academy of Sciences, Dolej\v{s}kova 3, 18223 Prague 8, Czech Republic}

\author{Jiří Brabec}
\affiliation{Heyrovsk\'{y} Institute, Czech Academy of Sciences, Dolej\v{s}kova 3, 18223 Prague 8, Czech Republic}

\author{Alexandra Tucholská}
\affiliation{Heyrovsk\'{y} Institute, Czech Academy of Sciences, Dolej\v{s}kova 3, 18223 Prague 8, Czech Republic}

\author{Mikuláš Matoušek}
\affiliation{Heyrovsk\'{y} Institute, Czech Academy of Sciences, Dolej\v{s}kova 3, 18223 Prague 8, Czech Republic}

\author{Katarzyna Pernal}
\affiliation{Institute of Physics, Lodz University of Technology, \mbox{ul.\ Wolczanska 217/221, 93-005 Lodz, Poland}}
\email{pernalk@gmail.com}

\author{Libor Veis}
\email{libor.veis@jh-inst.cas.cz}
\affiliation{Heyrovsk\'{y} Institute, Czech Academy of Sciences, Dolej\v{s}kova 3, 18223 Prague 8, Czech Republic}


\begin{document}

\maketitle

\begin{abstract}
Quantum hardware noise remains a major obstacle to accurate molecular simulations using the variational quantum eigensolver (VQE).
We introduce a novel error-mitigation strategy based on the extended random phase approximation (ERPA), which reconstructs the two-particle reduced density matrix (2-RDM) from approximate transition density matrices, effectively filtering errors in quantum measurements.
Unlike quantum subspace expansion, the approach requires only the one- and two-particle RDMs.
Since ERPA preserves the input 1-RDM, we further propose a hybrid quantum-classical scheme employing an inexpensive density matrix renormalization group (DMRG) calculation to provide a noise-free approximation to this quantity.
Benchmarks on water double-bond stretching and cyclobutadiene automerization, performed on IBM quantum hardware, demonstrate reductions in energy errors by orders of magnitude.
Combined with multireference adiabatic connection (AC0) corrections, the approach additionally recovers dynamical correlation beyond the active space, yielding accurate molecular potential energy curves and reaction barriers.
Our results establish ERPA as a promising, measurement-efficient route toward reliable quantum-classical electronic structure calculations.    
\end{abstract}

\section{Introduction}

Quantum computing offers a promising route toward the accurate simulation of strongly correlated molecules and materials, whose electronic structure remains challenging for classical computational methods~\cite{Cao2019,McArdle2020,HeadMarsden2020,Bauer2020,Motta2021}.
While quantum phase estimation (QPE)~\cite{aspuru-guzik_2005} provides
a systematic approach to obtaining molecular energies, its large circuit
depths necessitate quantum error correction~\cite{Reiher2017,Terhal2015},
placing practical applications beyond the capabilities of current quantum hardware.

For noisy intermediate-scale quantum (NISQ) devices~\cite{Preskill2018}, the variational quantum eigensolver (VQE)~\cite{Peruzzo2014,McClean2016,Tilly2022} has emerged as one of the most widely studied approaches.
By combining quantum circuits and energy measurements with classical parameter optimization, VQE reduces the quantum coherence requirements compared to QPE.
This makes it particularly attractive for near-term quantum chemistry applications, especially when employing compact, adaptively constructed ans\"atze such as ADAPT-VQE~\cite{Grimsley2019, Tang2021}.

Without quantum error correction, however, VQE calculations remain
susceptible to hardware noise arising from imperfect gate operations,
decoherence, and finite measurement statistics, which can introduce
substantial errors into calculated energies and molecular properties.
To address these limitations, numerous quantum error-mitigation
strategies have been developed~\cite{Cai2023},
including zero-noise extrapolation, probabilistic error cancellation,
symmetry verification, and quantum subspace expansion
(QSE)~\cite{McClean2017,Colless2018,McClean2020}.
These approaches aim to improve the accuracy of quantum computations
without the substantial overhead associated with full quantum error correction \cite{Katabarwa2024}.

Recently, we introduced a combination of VQE with the multireference adiabatic connection (AC) formalism \cite{Pernal2018, Pastorczak2018}, in which the one- and two-particle reduced density matrices (1-RDMs and 2-RDMs) obtained from a VQE calculation are used to recover dynamical electron correlation outside the complete active space (CAS) through classical post-processing. The resulting VQE-AC approach \cite{Matousek2024} provides a computationally affordable route toward a balanced description of static and dynamical correlation. 

In the present work, inspired by the relative insensitivity of AC corrections
to inaccuracies in the input RDMs \cite{Matousek2023}, we propose a novel
error-mitigation strategy based on the extended random phase approximation
(ERPA)~\cite{Chatterjee2012}.

By reconstructing the ground-state 2-RDM through a spectral
decomposition based on noisy 1- and 2-RDMs measured on a quantum
computer, ERPA provides a means of suppressing errors in the
two-particle quantities.
The proposed approach shares certain features with quantum subspace
expansion (QSE), but requires substantially fewer measurements,
involving only the 1- and 2-RDMs.
Since the ERPA reconstruction preserves the input 1-RDM, we also
investigate a hybrid strategy in which the noisy quantum-measured
1-RDM is replaced by an approximate, noise-free 1-RDM obtained
from an inexpensive density matrix renormalization group (DMRG)
calculation~\cite{chan_review,reiher_perspective}.

Benchmark calculations for the symmetric double-bond stretching
of water and the automerization of cyclobutadiene demonstrate
that the proposed strategy can reduce errors in quantum-hardware
energies by orders of magnitude.

The remainder of this paper is organized as follows.
We first present the theoretical framework, including VQE, QSE, ERPA-based
error mitigation, and AC.
After describing the computational details, we assess the proposed methodology
using both finite-shot state vector simulations and calculations on IBM quantum hardware.
Finally, we summarize our findings and discuss the prospects of the approach
for quantum-classical electronic structure calculations.

\section{Theory}

\subsection{Variational Quantum Eigensolver}

The variational quantum eigensolver (VQE)~\cite{Peruzzo2014, McClean2016} is a hybrid quantum-classical algorithm in which a parameterized quantum circuit prepares a trial state $|\Psi(\boldsymbol{\theta})\rangle$ and the parameters $\boldsymbol{\theta}$ are optimized by minimizing the energy
\begin{equation}
E_{\mathrm{VQE}}
=
\min_{\boldsymbol{\theta}}
\langle\Psi(\boldsymbol{\theta})|
\hat{H}
|\Psi(\boldsymbol{\theta})\rangle .
\label{eq:vqe_energy}
\end{equation}
For the electronic Hamiltonian in second quantization,
\begin{equation}
  \hat{H} = \sum_{pq} h_{pq} \hat{a}_{p}^{\dagger} \hat{a}_{q} + \frac{1}{2} \sum_{pqrs} \langle pq | rs \rangle \hat{a}_{p}^{\dagger} \hat{a}_{q}^{\dagger} \hat{a}_{s} \hat{a}_{r} ,
  \label{ham_sec_quant}
\end{equation}
where $h_{pq}$ and $\langle pq|rs\rangle$ denote the one- and two-electron integrals in a molecular spin-orbital basis~\cite{szabo_ostlund}, respectively. Fermionic operators are transformed into qubit operators using, e.g., the Jordan--Wigner~\cite{jordan_1928, whitfield_2010} or Bravyi--Kitaev~\cite{Bravyi2002, Seeley2012} mappings, allowing the energy to be evaluated from measurements of the corresponding Pauli strings on the quantum-prepared state.

Since $\hat H$ contains at most two-body interactions, its expectation value is completely determined by the one- and two-particle reduced density matrices (RDMs). 

\begin{eqnarray}
  \gamma_{pq}(\boldsymbol{\theta}) & = & \bra{\Psi(\boldsymbol{\theta})} \hat{a}_{p}^{\dagger} \hat{a}_{q} \ket{\Psi(\boldsymbol{\theta})} \label{g1} \\
  \Gamma_{pqrs}(\boldsymbol{\theta}) & = & \bra{\Psi(\boldsymbol{\theta})} \hat{a}_{p}^{\dagger} \hat{a}_{q}^{\dagger} \hat{a}_{s} \hat{a}_{r} \ket{\Psi(\boldsymbol{\theta})} \label{g2}
\end{eqnarray}

\noindent
These quantities therefore provide a natural interface between the VQE calculation and the RDM-based error-mitigation approaches considered below.

In the present work, the VQE trial states are constructed using the qubit-ADAPT-VQE approach~\cite{Tang2021}, a qubit-space variant of ADAPT-VQE~\cite{Grimsley2019}. In contrast to fixed ans\"atze, ADAPT-VQE iteratively constructs the parametrized circuit by selecting operators from a predefined pool according to their contribution to lowering the energy. Starting from a reference state $|\Psi_{\mathrm{ref}}\rangle$, the wave function after $N$ adaptive iterations can be written as
\begin{equation}
|\Psi(\boldsymbol{\theta})\rangle
=
\prod_{k=1}^{N}
e^{\theta_k \hat{P}_k}
|\Psi_{\mathrm{ref}}\rangle ,
\label{eq:adapt_ansatz}
\end{equation}
where $\hat{P}_k$ denotes the operator selected at the $k$th iteration and $\theta_k$ is the corresponding variational parameter. At each iteration, the operator yielding the largest energy gradient is appended to the ansatz, followed by reoptimization of all variational parameters.

While the original ADAPT-VQE employs a pool of fermionic excitation operators, qubit-ADAPT-VQE constructs the operator pool directly from Pauli strings~\cite{Tang2021}, building on qubit-space coupled-cluster parametrizations~\cite{Ryabinkin2018}. This eliminates the need to implement complete mapped fermionic excitation operators and can substantially reduce the circuit depth. Qubit-ADAPT-VQE is therefore adopted here to obtain compact approximations to the molecular ground states used for subsequent RDM measurements and error mitigation.

\subsection{Quantum Subspace Expansion}

The quantum subspace expansion (QSE) \cite{McClean2017, Colless2018, McClean2020}, as originally proposed, provides a systematic framework for mitigating errors in VQE calculations by variationally diagonalizing the Hamiltonian within a subspace generated by a selected set of excitation operators acting on the VQE reference state. The resulting generalized eigenvalue problem can be constructed from reduced quantities measured on the quantum device, enabling improved estimates of energies and other observables. Beyond ground-state error mitigation, QSE can also be employed to access excited states, although the resulting excitation energies are generally not size-intensive \cite{Asthana2023}. 

QSE has further been extended to recover dynamical electron correlation outside an active space \cite{Takeshita2020}. However, as discussed below, its construction generally involves reduced density matrices (RDMs) of order higher than two, substantially increasing the quantum measurement cost. Higher-order RDM measurements can be avoided through cumulant approximations, although their accuracy may deteriorate in strongly correlated regimes, precisely where quantum computers are expected to offer the greatest advantage over classical methods.

Let $|\Psi_0\rangle$ denote the optimized VQE state, where, for clarity, we suppress its explicit dependence on the ansatz parameters $\boldsymbol{\theta}$, and let ${\hat{O}_\mu}$ denote a set of operators defining the expansion manifold, such as single and double excitation operators. The corresponding QSE basis states are

\begin{equation}
|\Phi_\mu\rangle = \hat O_\mu |\Psi_0\rangle .
\end{equation}
Any eigenstate can be expanded in this basis
\begin{equation}
|\Psi_n^{\mathrm{QSE}}\rangle
= \sum_\mu c_{\mu n} |\Phi_\mu\rangle =
\sum_\mu c_{\mu n}\hat O_\mu|\Psi_0\rangle .
\label{eq:qse_ansatz}
\end{equation}

\noindent
Application of the Rayleigh--Ritz variational principle within this generally nonorthogonal subspace leads to the generalized eigenvalue problem
\begin{equation}
\mathbf H^{\mathrm{QSE}}\mathbf c_n
=
E_n^{\mathrm{QSE}}
\mathbf S^{\mathrm{QSE}}\mathbf c_n ,
\label{eq:qse_generalized_evp}
\end{equation}
where

\begin{eqnarray}
H_{\mu\nu}^{\mathrm{QSE}}
& = &
\langle\Psi_0|
\hat O_\mu^\dagger
\hat H
\hat O_\nu
|\Psi_0\rangle \label{eq:qse_hamiltonian_matrix} \\
S_{\mu\nu}^{\mathrm{QSE}}
& = &
\langle\Psi_0|
\hat O_\mu^\dagger
\hat O_\nu
|\Psi_0\rangle .
\label{eq:qse_overlap_matrix}
\end{eqnarray}

\noindent
The lowest eigenvalue yields the QSE estimate of the ground-state energy within the chosen subspace, while the higher eigenvalues provide approximations to the corresponding excited-state energies.

In the present context, it is particularly useful to consider a QSE space generated by generalized one-body excitation operators,
\begin{equation}
\hat E_{pq}=a_p^\dagger a_q,
\label{eq:qse_single_operator}
\end{equation}
where $p$ and $q$ denote general spin-orbital indices. The operator manifold is also augmented by the identity operator to ensure that the original VQE reference state is explicitly included in the QSE subspace. The Hamiltonian and overlap matrices then take the form
\begin{eqnarray}
H_{pq,rs}^{\mathrm{QSE}}
& = &
\langle\Psi_0|
a_q^\dagger a_p
\hat H
a_r^\dagger a_s
|\Psi_0\rangle .
\label{eq:qse_hamiltonian_singles} \\
S_{pq,rs}^{\mathrm{QSE}}
& = &
\langle\Psi_0|
a_q^\dagger a_p
a_r^\dagger a_s
|\Psi_0\rangle .
\label{eq:qse_overlap_singles}
\end{eqnarray}

Using the fermionic anticommutation relations,
\begin{equation}
\{a_p,a_q^\dagger\}=\delta_{pq},
\end{equation}
the overlap matrix in Eq.~\eqref{eq:qse_overlap_singles} can be expressed in terms of the one- and two-particle RDMs defined in Eqs. \eqref{g1} and \eqref{g2}.
In contrast, for a two-body electronic Hamiltonian \eqref{ham_sec_quant}, the matrix elements $H_{pq,rs}^{\mathrm{QSE}}$ generally involve RDMs up to fourth order. Thus, the generalized singles QSE requires many-body information beyond the 2-RDM, substantially increasing the associated measurement cost.

The error-mitigation capability of QSE originates from embedding the approximate VQE reference state in a larger, physically motivated variational subspace,
\begin{equation}
\mathcal K_{\mathrm{QSE}}
=
\mathrm{span}\{
\hat O_\mu|\Psi_0\rangle
\},
\end{equation}
which explicitly contains the original VQE state, $|\Psi_0\rangle\in\mathcal K_{\mathrm{QSE}}$. Variational diagonalization of the Hamiltonian within this enlarged subspace yields the QSE ground-state approximation
\begin{equation}
|\Psi_0^{\mathrm{QSE}}\rangle
=
\sum_\mu c_{\mu0}\hat O_\mu|\Psi_0\rangle .
\end{equation}
The additional directions, $\hat O_\mu|\Psi_0\rangle$, allow QSE to correct components of the reference-state error that lie within the chosen operator-generated subspace. For exact QSE matrix elements, the Rayleigh--Ritz variational principle consequently yields
\begin{equation}
E_{\mathrm{exact}}
\leq
E_0^{\mathrm{QSE}}
\leq
E_{\mathrm{VQE}} .
\end{equation}

QSE also provides a useful point of comparison with the extended random phase approximation (ERPA) \cite{Chatterjee2012, Pernal2017} considered in the following section. Both approaches construct a generalized eigenvalue problem from expectation values evaluated with respect to an approximate reference state and employ an operator-generated manifold to describe many-body fluctuations around that state. Their underlying formulations, however, are different. QSE is a variational state-space method based on Hamiltonian matrix elements \eqref{eq:qse_hamiltonian_matrix},
whereas ERPA follows from an equation-of-motion formulation involving commutators of the Hamiltonian with the excitation operators. 
For generalized one-body excitation operators and a two-body Hamiltonian, this commutator structure allows the ERPA equations to be expressed entirely in terms of the one- and two-particle RDMs \cite{Pernal2017}.

This structural similarity motivates the use of ERPA as an alternative post-processing approach to error mitigation. Rather than variationally improving the reference state, as in QSE, ERPA uses the measured VQE RDMs to construct an approximate many-body response and, as described in the following section, to reconstruct the ground-state 2-RDM from the resulting transition densities. 
%\lv{The ERPA reconstruction may thereby correct errors in the measured RDM that are inconsistent with the response structure imposed by the ERPA equations.}

Although ERPA also has a variational formulation \cite{Ollitrault2020}, the resulting method is not generally variational, since the killer condition \eqref{killer_condition} need not hold and the reference state need not be the exact ground state. Nevertheless, its effectiveness in error mitigation, assessed numerically in this work, may be partly attributed to its variational foundation.

\subsection{Extended Random Phase Approximation}

A central relation underlying the RDM formulation of ERPA is the McLachlan--Ball relation, which expresses the ground-state 2-RDM in terms of the ground-state 1-RDM and one-particle transition density matrices \cite{McLachlan1964,Pernal2017}. Using the generalized one-body operators introduced above, the fermionic anticommutation relations give
\begin{equation}
a_p^\dagger a_q^\dagger a_s a_r
=
\hat E_{pr}\hat E_{qs}
-
\delta_{rq}\hat E_{ps}.
\end{equation}
Inserting the complete resolution of identity,
\begin{equation}
\hat I
=
|\Psi_0\rangle\langle\Psi_0|
+
\sum_{n\neq0}
|\Psi_n\rangle\langle\Psi_n|,
\end{equation}
between the one-body operators yields
\begin{equation}
\Gamma_{pqrs}
=
\gamma_{pr}\gamma_{qs}
+
\sum_{n\neq0}
\gamma_{pr}^{0n}\gamma_{qs}^{n0}
-
\delta_{rq}\gamma_{ps},
\label{eq:mclachlan_ball}
\end{equation}
where
\begin{eqnarray}
\gamma_{pq}^{0n}
& = &
\langle\Psi_0|
\hat E_{pq}
|\Psi_n\rangle, \\
\gamma_{pq}^{n0}
& = &
\langle\Psi_n|
\hat E_{pq}
|\Psi_0\rangle
\end{eqnarray}
are the one-particle transition density matrices between the ground and excited states. Equation~\eqref{eq:mclachlan_ball} is exact when the sum extends over a complete set of many-body eigenstates.

In practice, the exact excited states and the corresponding transition density matrices entering Eq.~\eqref{eq:mclachlan_ball} are not available. ERPA provides an approximate route to these transition density matrices, together with the associated excitation energies, through an equation-of-motion (EOM) formulation. For an exact excited state $|\Psi_n\rangle$, let $\hat R_n^\dagger$ denote an excitation operator satisfying
\begin{equation}
|\Psi_n\rangle
=
\hat R_n^\dagger |\Psi_0\rangle .
\end{equation}
Using the Schr\"odinger equations
\begin{equation}
\hat H|\Psi_0\rangle = E_0|\Psi_0\rangle,
\qquad
\hat H|\Psi_n\rangle = E_n|\Psi_n\rangle,
\end{equation}
one obtains
\begin{equation}
[\hat H,\hat R_n^\dagger]|\Psi_0\rangle
=
\omega_n \hat R_n^\dagger|\Psi_0\rangle ,
\label{eq:eom_exact}
\end{equation}
where $\omega_n=E_n-E_0$ denotes the excitation energy.

Projecting Eq.~\eqref{eq:eom_exact} onto the excited-state manifold using the
corresponding de-excitation operators $\hat R_m$ and invoking the ground-state
annihilation condition (so called killer condition)
\begin{equation}
    \hat R_m|\Psi_0\rangle = 0 ,
    \label{killer_condition}
\end{equation}
the EOM can be written in the double-commutator form
\begin{equation}
    \langle\Psi_0|
    [\hat R_m,[\hat H,\hat R_n^\dagger]]
    |\Psi_0\rangle
    =
    \omega_n
    \langle\Psi_0|
    [\hat R_m,\hat R_n^\dagger]
    |\Psi_0\rangle .
    \label{eq:eom_double_commutator}
\end{equation}

Within ERPA, the excitation operator is approximated by a linear combination of generalized one-body operators containing forward and backward components,
\begin{equation}
\hat R_n^\dagger
=
\sum_{r>s}
\left(
X_{rs}^{n}\hat E_{rs}
-
Y_{rs}^{n}\hat E_{sr}
\right),
\label{eq:erpa_excitation_operator}
\end{equation}
where $X_{rs}^{n}$ and $Y_{rs}^{n}$ denote the corresponding forward and backward amplitudes, respectively. 

For the two-body Hamiltonian in Eq.~\eqref{ham_sec_quant}, the commutator of $\hat H$ with a one-body operator $\hat E_{rs}$ contains operators of at most two-body rank. The subsequent commutation with another one-body operator does not increase this rank. Consequently, all ground-state expectation values entering the double-commutator EOM can be evaluated entirely in terms of the 1- and 2-RDMs.

Substitution of Eq.~\eqref{eq:erpa_excitation_operator} and the corresponding de-excitation operator into Eq.~\eqref{eq:eom_double_commutator} yields the ERPA generalized eigenequation
\begin{equation}
    \begin{pmatrix}
        \mathbf A & \mathbf B \\
        \mathbf B & \mathbf A
    \end{pmatrix}
    \begin{pmatrix}
        \mathbf X^{n} \\
        \mathbf Y^{n}
    \end{pmatrix}
    =
    \omega_n
    \begin{pmatrix}
        -\mathbf N & \mathbf 0 \\
        \mathbf 0 & \mathbf N
    \end{pmatrix}
    \begin{pmatrix}
        \mathbf X^{n} \\
        \mathbf Y^{n}
    \end{pmatrix},
    \label{eq:erpa_evp}
\end{equation}
where $\mathbf A$ and $\mathbf B$ arise from the double-commutator matrix elements and $\mathbf N$ is the corresponding metric matrix obtained from single commutators. Their explicit expressions can be written solely in terms of the one- and two-electron integrals and the ground-state 1- and 2-RDMs and can be found e.g. in Ref.~\cite{Pernal2017}.

The one-particle transition density matrix associated with the $n$th ERPA mode is
\begin{equation}
    \gamma_{pq}^{0n}
    =
    \langle\Psi_0|
    \hat E_{pq}
    |\Psi_n\rangle
    =
    \langle\Psi_0|
    \hat E_{pq}\hat R_n^\dagger
    |\Psi_0\rangle .
\end{equation}
Using the adjoint form of the killer condition,
\begin{equation}
    \langle\Psi_0|\hat R_n^\dagger = 0,
\end{equation}
the transition density can equivalently be written as
\begin{equation}
    \gamma_{pq}^{0n}
    =
    \langle\Psi_0|
    [\hat E_{pq},\hat R_n^\dagger]
    |\Psi_0\rangle .
    \label{eq:erpa_trdm_commutator}
\end{equation}
Substituting the ERPA excitation operator from
Eq.~\eqref{eq:erpa_excitation_operator} into
Eq.~\eqref{eq:erpa_trdm_commutator}, and using
\begin{equation}
    [\hat E_{pq},\hat E_{rs}]
    =
    \delta_{qr}\hat E_{ps}
    -
    \delta_{sp}\hat E_{rq},
\end{equation}
one obtains
\begin{equation}
\begin{split}
    \gamma_{pq}^{0n,\mathrm{ERPA}}
    ={}&
    \sum_{s<q} X_{qs}^{n}\gamma_{ps}
    -
    \sum_{r>p} X_{rp}^{n}\gamma_{rq}
    \\
    &-
    \sum_{r>q} Y_{rq}^{n}\gamma_{pr}
    +
    \sum_{s<p} Y_{ps}^{n}\gamma_{sq}.
\end{split}
\label{eq:erpa_transition_density}
\end{equation}

Motivated by the error-mitigation approach discussed above in the context of
QSE, we now consider the ERPA transition density matrices as a means of
reconstructing the VQE 2-RDM through Eq.~\eqref{eq:mclachlan_ball}. An
important consistency property of this reconstruction is that its partial
contraction recovers the input 1-RDM. To show this, we first note that the
ERPA transition density matrices are traceless. Indeed,

\begin{equation}
    \sum_p \gamma_{pp}^{0n,\mathrm{ERPA}}
    =
    \langle\Psi_0|
    [\hat N,\hat R_n^\dagger]
    |\Psi_0\rangle
    =0,
\end{equation}
where $\hat N=\sum_p\hat E_{pp}$ and
$[\hat N,\hat R_n^\dagger]=0$, because the ERPA excitation operator conserves
particle number. Consequently,

\begin{eqnarray}
    \sum_q \Gamma_{pqrq}^{\mathrm{ERPA}}
    & = &
    \gamma_{pr}\sum_q\gamma_{qq} \nonumber \\
    & + & \sum_{n\neq0}
    \gamma_{pr}^{0n,\mathrm{ERPA}}
    \sum_q\gamma_{qq}^{n0,\mathrm{ERPA}} \nonumber \\
    & - &
    \sum_q\delta_{rq}\gamma_{pq}
    \nonumber\\
    &= &
    (N-1)\gamma_{pr}.
\end{eqnarray}

\noindent
Thus, contraction of the reconstructed 2-RDM yields exactly the input 1-RDM.
Consequently, ERPA modifies the two-particle density matrix while leaving its
associated one-particle density matrix unchanged and therefore cannot, by
itself, mitigate errors already present in the input 1-RDM. This constitutes
a limitation of the ERPA reconstruction. On the other hand, an important
advantage over QSE is its lower measurement requirement since the ERPA formulation
considered here requires quantum measurements only up to the two-particle
level, i.e., the 1- and 2-RDMs.

The inability of ERPA to refine the input 1-RDM may be partially alleviated
by allocating a larger fraction of a finite measurement budget to its
estimation. Since the 1-RDM contains substantially fewer elements than the
2-RDM, it can be measured with higher statistical accuracy for a fixed total
measurement budget, thereby reducing the noise carried by the 1-RDM entering
the ERPA reconstruction \cite{LewYee2025,Tilly2021}.

A further possibility is to replace the measured 1-RDM with a noise-free
classical approximation, for example from a low-bond-dimension DMRG
calculation \cite{White-1992, White-1993, chan_review, reiher_perspective}. Since the ERPA-reconstructed 2-RDM contracts exactly to the
supplied 1-RDM, this may provide an additional source of error mitigation.
The ERPA response equations constrain the two-particle structure, while the
contraction relation tailors the reconstruction to the supplied noise-free
1-RDM. The latter is, however, only an approximation to the exact 1-RDM, and
the effectiveness of this strategy is examined numerically in the Results
section.

\subsection{Adiabatic Connection}

The techniques discussed in the preceding sections aim to improve the VQE
description within the complete active space (CAS), which is chosen to
contain the orbitals responsible for strong electron correlation. Accordingly,
all orbital indices in $\hat{H}$ are restricted to the active space, while
the one-electron integrals include the frozen-core contribution from the
doubly occupied orbitals below the CAS. In our framework, the predominantly
dynamical correlation involving orbitals outside the CAS is accounted for
using the multireference adiabatic connection (AC) method
\cite{Pernal2018,Pastorczak2018,Pastorczak2019}.

Within this framework, the total energy is written as an additive correction
to the CAS energy,
\begin{equation}
    E = E_{\mathrm{CAS}} + E_{\mathrm{corr}}^{\mathrm{AC}},
\end{equation}
where
$E_{\mathrm{CAS}} =
\langle \Psi_{\mathrm{CAS}} | \hat{H} | \Psi_{\mathrm{CAS}} \rangle$,
and $E_{\mathrm{corr}}^{\mathrm{AC}}$ accounts for correlation effects
missing from the active-space description. The additive character of AC is
particularly important in the present context. AC recovers the missing
out-of-CAS correlation but does not modify the underlying CAS energy.
Consequently, errors in the VQE estimate of $E_{\mathrm{CAS}}$ propagate
directly to the final AC-corrected energy, motivating the error-mitigation
procedure introduced in the preceding section.

The AC construction introduces a linear connection between the zeroth-order
Hamiltonian, $\hat{H}^{(0)}$, and the full-space electronic Hamiltonian,
$\hat{H}$ of Eq.~\eqref{ham_sec_quant}, with orbital indices now spanning
the complete orbital space.
\begin{eqnarray}
  \hat{H}^{\alpha} = \hat{H}^{(0)} + \alpha \hat{H}^{\prime},
  \quad \text{with} \quad
  \hat{H}^{\prime} = \hat{H} - \hat{H}^{(0)}
  \label{h_alpha} \\
  \text{and} \quad \alpha:0 \rightarrow 1 .
  \nonumber
\end{eqnarray}
\noindent
For a CAS reference state, $\hat{H}^{(0)}$ may be chosen as either the
group product function Hamiltonian~\cite{Pernal2018,Pastorczak2018} or
the Dyall Hamiltonian~\cite{Matousek2023}.

Application of the Hellmann--Feynman theorem yields the exact AC
expression for the correlation energy,
\begin{eqnarray}
  E^{\text{AC}}_{\text{corr}} & = & \int_0^1 W^{\alpha} \rm{d}\alpha
  \label{ac_corr} \\
  W^{\alpha} & = &
  \bra{\Psi^{\alpha}} \hat{H}^{\prime} \ket{\Psi^{\alpha}}\nonumber \\
  & - &
  \bra{\Psi_{\text{CAS}}} \hat{H}^{\prime}
  \ket{\Psi_{\text{CAS}}},
  \label{W}
\end{eqnarray}
\noindent
where $W^{\alpha}$ denotes the exact AC integrand and
$\Psi^{\alpha}$ is the ground state of $\hat{H}^{\alpha}$
[Eq.~(\ref{h_alpha})]. Direct evaluation of this exact expression is
not practical, and several approximations are therefore introduced to
obtain a computationally applicable form of Eq.~(\ref{ac_corr})
~\cite{Pernal2018,Pastorczak2018}.

A central assumption of approximate AC schemes is that the 1-RDM,
$\gamma$, remains unchanged along the AC path, commonly referred to as
the fixed-RDM approximation~\cite{Matousek2023}. This assumption is
motivated by the predominantly dynamical character of the electron
correlation recovered by AC, which is expected to alter 1-RDM only marginally. 
Furthermore, the $\alpha$-dependent 2-RDMs
entering Eq.~\eqref{W} are expressed through the McLachlan--Ball
relation, Eq.~\eqref{eq:mclachlan_ball}, in terms of
$\alpha$-dependent transition density matrices, which are subsequently
approximated within ERPA~\cite{Chatterjee2012}. The ERPA equations were
introduced in the preceding section, while explicit expressions for
the $\alpha$-dependent ERPA matrices can be found, for example, in
Ref.~\cite{Pastorczak2018}.

The linearized-AC-integrand approximation, AC0, avoids the numerical
integration in Eq.~\eqref{ac_corr} while retaining most of the accuracy
of the full AC treatment, as demonstrated numerically in
Refs.~\cite{Pastorczak2018,Pastorczak2019}.

Since the AC correlation energy is evaluated by solving the
$\alpha$-dependent ERPA equations, it is inherently relatively
insensitive to the quality of the input RDMs. This robustness was
demonstrated previously in a classical setting~\cite{Matousek2023}
and will be examined here for RDMs obtained from VQE calculations.

\section{Computational details}

\begin{figure}[!ht]
  \includegraphics[width=0.9\linewidth]{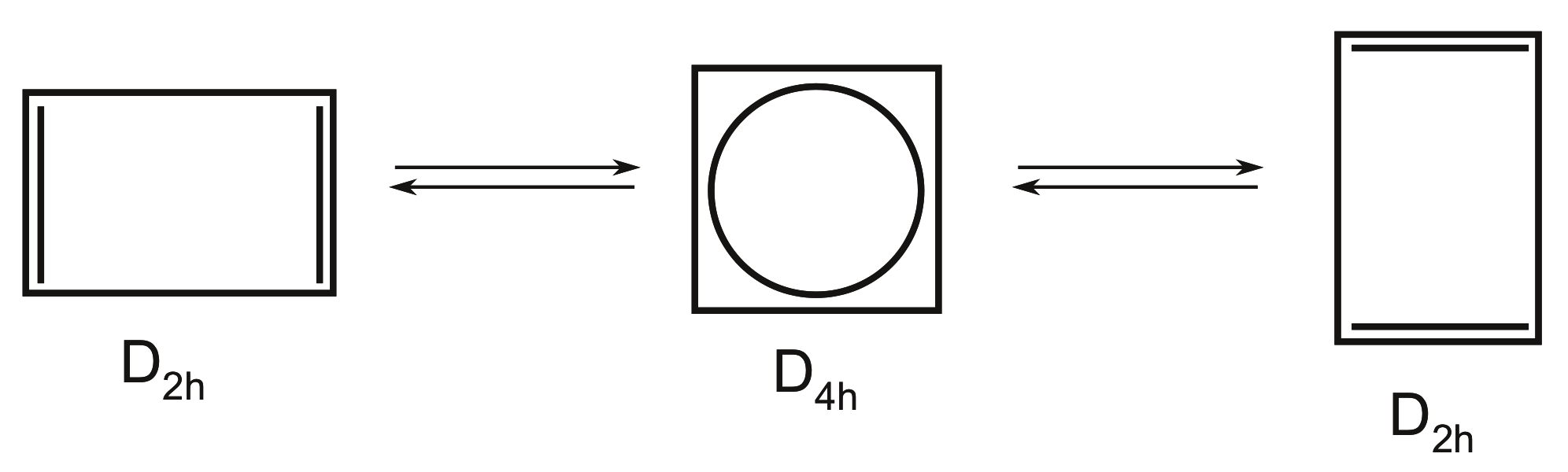}
  \caption{The automerization reaction of cyclobutadiene, with the corresponding molecular point groups indicated.}
  \label{cb_reaction}
\end{figure}

The ERPA error-mitigation scheme combined with the AC0 correction was
tested on two benchmark problems characterized by strong electron
correlation, namely the symmetric double-bond stretching of the water molecule ($\mathrm{H_2O}$, $\angle\mathrm{H\!-\!O\!-\!H} = 107.5^\circ$)
and the automerization of cyclobutadiene, as depicted in Figure \ref{cb_reaction}.

Correlation-consistent basis sets \cite{Dunning1989} were employed for both systems, specifically cc-pVDZ for water and cc-pVTZ for cyclobutadiene. The geometries along the cyclobutadiene automerization pathway were taken from Ref.~\cite{EckertMaksi2006}. For both molecules, the active space comprised four electrons in four orbitals, CAS(4,4), corresponding to the orbitals involved in bond breaking in water and the $\pi$ orbitals in cyclobutadiene. The molecular orbitals were obtained from the respective CASSCF(4,4) calculations. Accordingly, the bare VQE energies are expected to approach the corresponding CASSCF values.

To disentangle the effects of noise on energy estimation from parameter
optimization and adaptive ans\"atz construction, we adopted the following
computational protocol. First, the wave-function ansatz and its parameters
were optimized using qubit-ADAPT-VQE \cite{Tang2021} with the state-vector
simulator in Qiskit \cite{qiskit2024}. Unlike fermionic excitation-based
ans\"atze, the qubit-ADAPT ansatz does not preserve particle-number and
spin symmetries by construction. Nevertheless, we employed the qubit
formulation because it enables compact ans\"atze and shallow quantum
circuits, which are advantageous for implementation on current quantum
hardware. While no observable symmetry breaking occurred for water,
a spin-penalty term was added to the Hamiltonian for cyclobutadiene
to prevent spin-symmetry breaking during optimization. We verified
that the optimized states retained the correct particle number and
spin symmetry. 

The qubit-ADAPT-VQE ansatz was constructed using 10 adaptive iterations for water and 15 for cyclobutadiene, with one operator added at each iteration, followed by reoptimization of all variational parameters. This ensured that the noise-free VQE energies deviated from the corresponding CASSCF(4,4) references by less than 1 m$E_h$. After transpilation, the resulting circuits contained approximately 80 two-qubit gates.

The optimized ans\"atz and its parameters were then kept fixed and
transferred to an IBM quantum processing unit (QPU), where the
expectation values required to reconstruct the symmetry-nonredundant
elements of the 1- and 2-RDMs were measured. Thus, ansatz construction
and parameter optimization were performed in a noise-free setting,
whereas hardware noise affected only the QPU evaluation of the RDMs
and the corresponding energy.

The resulting RDMs were subsequently used to compute the
ERPA-mitigated CAS energy and the AC0 correlation correction, providing
the final estimate of the total energy.

The quantum-hardware calculations were performed on the 156-qubit
\texttt{ibm\_pittsburgh} QPU, based on the Heron r3 architecture.

The parity fermion-to-qubit mapping \cite{Seeley2012} was employed,
followed by symmetry-based qubit tapering \cite{Bravyi_2017, Setia2020}.
This reduced the number of qubits required for the CAS(4,4)
calculations to five for H$_2$O and four for cyclobutadiene.
 To make more efficient use of the
available QPU resources, multiple instances of each measurement circuit
were executed in parallel on distinct subsets of physical qubits.
These subsets were selected using backend calibration data, taking into
account readout and gate errors, gate durations, connectivity, and the
two-qubit interactions required by the logical circuit. Candidate
logical-to-physical mappings were trial-transpiled at Qiskit optimization
level~3 and ranked according to circuit depth, native two-qubit gate
count, physical distances between interacting qubits, and accumulated
gate and readout error penalties. Mutually disjoint regions were then
selected while additionally discouraging neighboring placements in order to avoid cross-talks. Subject
to these calibration and routing constraints, 7--10 circuit replicas
could be executed simultaneously,
depending on the molecular geometry. Each physical shot measured all
replicas in the same measurement basis simultaneously, thereby reducing
the required number of physical shot cycles by corresponding factors
relative to serial execution at matched aggregate sample counts. Ten
independent QPU runs were performed to assess the statistical variation
of the resulting RDMs and energies.

All CASSCF, DMRG, and ERPA/AC0 calculations were performed using
\textsf{ORCA} \cite{orca}, \textsf{MOLMPS} \cite{Brabec2020, molmps_scalable}, and \textsf{GAMMCOR} \cite{gammcor}, respectively.

\section{Results}
\subsection{Symmetric stretching of water molecule}

We first investigate the effect of finite-sampling (shot) noise in a
classical simulation, prior to considering actual quantum hardware.
An important aspect of the present ERPA-based error-mitigation scheme
is that the input 1-RDM is not altered by ERPA and
should therefore be determined with the highest possible accuracy. In
practice, this motivates allocating a larger number of measurements,
or potentially more advanced error-mitigation techniques (in case of hardware noise), to the
1-RDM than to the 2-RDM. 

\begin{figure*}[!ht]
    \centering
    \subfloat[\label{h2o_shots_1000}]{%
    \includegraphics[width=0.5\linewidth]{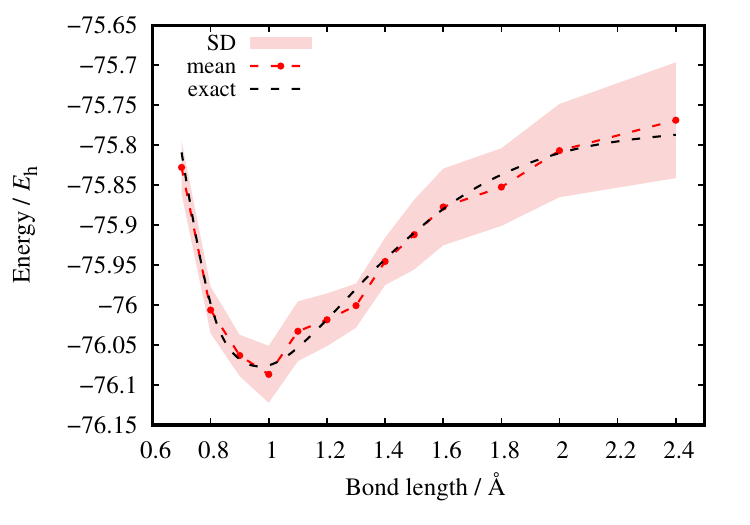}
    }
    \subfloat[\label{h2o_shots_10000}]{%
    \includegraphics[width=0.5\linewidth]{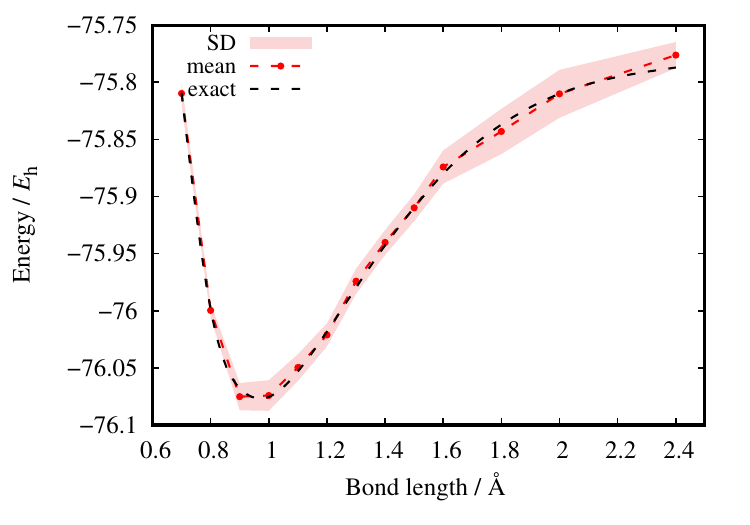}
    } \\
    \subfloat[\label{h2o_shots_1000_corr}]{%
    \includegraphics[width=0.5\linewidth]{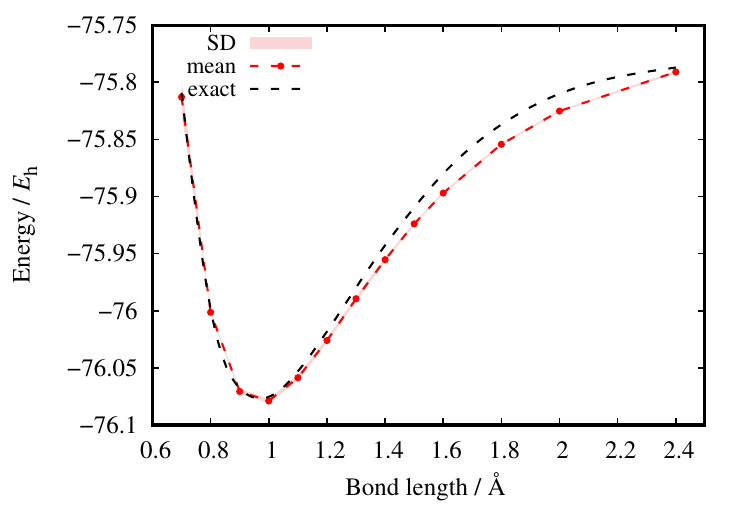}
    }
    \subfloat[\label{h2o_shots_10000_corr}]{%
    \includegraphics[width=0.5\linewidth]{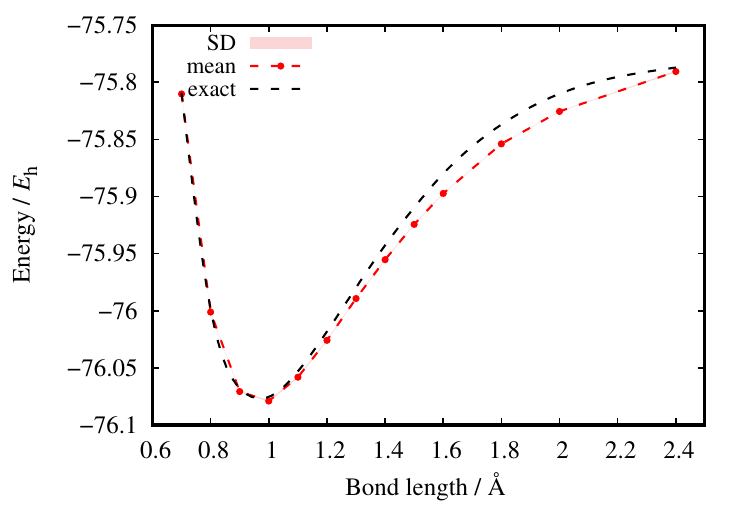}
    }    
    \caption{Potential energy curves for the symmetric stretching of H$_2$O obtained
from finite-shot evaluation of the optimized VQE state in the CAS(4,4)
active space using Qiskit Aer. The 1-RDMs were taken to be exact, while
the 2-RDMs were evaluated with finite-shot sampling. Red points and shaded
regions represent the mean and $\pm 1$ standard deviation (SD), respectively,
over 10 independent runs. The dashed black line denotes the CASSCF(4,4)
reference. (a,b) Energies obtained with 1,000 and 10,000 shots,
respectively; (c,d) corresponding ERPA-corrected energies.}
    \label{h2o_shot_noise}
\end{figure*}

To isolate the effect of sampling noise in the
2-RDM, Fig.~\ref{h2o_shot_noise} shows results obtained with an exact
1-RDM and shot-noise-affected 2-RDMs. In the absence of other noise
sources, the exact 1-RDM represents the limiting case of sufficiently
high sampling accuracy. As expected, increasing the number of shots
from 1,000 to 10,000 substantially reduces the statistical uncertainty
of the resulting energies, consistent with the expected
$1/\sqrt{N_{\mathrm{shots}}}$ scaling of sampling errors.

The corresponding ERPA-corrected energies are shown in
Figs.~\ref{h2o_shots_1000_corr} and \ref{h2o_shots_10000_corr}.
ERPA efficiently suppresses the fluctuations originating from shot
noise in the input 2-RDM by reconstructing it under the constraint of
reproducing the noise-free 1-RDM. Remarkably, the statistical
uncertainty of the ERPA-corrected energies is very similar for 1,000
and 10,000 shots, indicating that the ERPA reconstruction largely
eliminates the dependence on the sampling accuracy of the input 2-RDM
in this regime. This substantial reduction of the statistical
uncertainty, however, comes at the cost of a systematic
bias in the reconstructed CAS energies. The effect is most pronounced
for $r=1.5-2.0$~\AA, where the ERPA-corrected energies fall below
the exact values, with a maximum deviation of approximately
$15$~m$E_h$ at $r=1.8$~\AA.

The occurrence of energies below the exact ground-state energy reflects
the fact that the ERPA-reconstructed 2-RDM is not guaranteed to be
$N$-representable. Such a bias is intrinsic to the present ERPA
reconstruction. Even when exact (FCI) RDMs are supplied as input,
the truncated excitation-operator expansion in
Eq.~\eqref{eq:erpa_excitation_operator} does not, in general, reconstruct
the exact 2-RDM. Indeed, for H$_2$/cc-pVTZ with FCI input RDMs, ERPA
reconstruction was previously shown to introduce negative energy errors of up to
approximately $5$~m$E_h$~\cite{Pernal2017}. In principle, the
reconstructed 2-RDM could be further purified to approximately restore
$N$-representability, for example using the techniques of
Ref.~\cite{Lanssens2018}, nevertheless such an additional purification step is not
considered here.

It is also worth noting that this negative bias may be partially
compensated in actual QPU calculations by errors arising from imperfect
state preparation, including gate errors and decoherence. Such
imperfections generally deteriorate the prepared ground-state
approximation and are therefore expected to increase its variational
energy. Their interplay with the negative bias introduced by the ERPA
reconstruction will be examined below for the hardware calculations.

\begin{figure*}[!ht]
    \centering
    \subfloat[\label{h2o_shots_1000_qpu}]{%
    \includegraphics[width=0.5\linewidth]{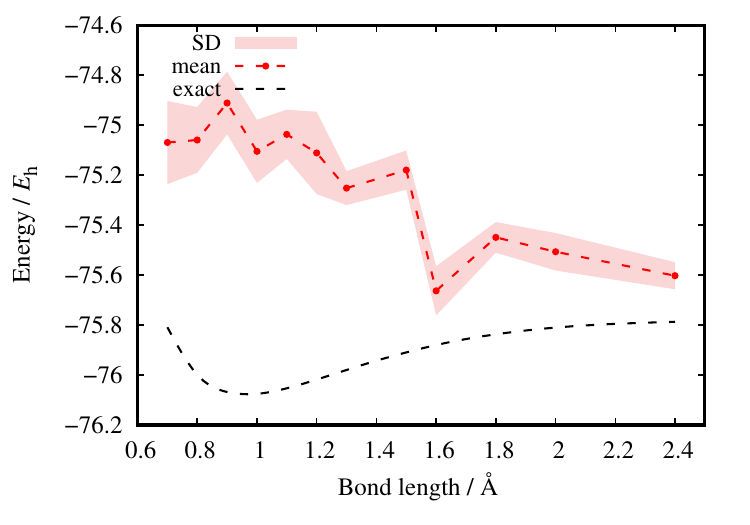}
    }
    \subfloat[\label{h2o_shots_1000_qpu_corr}]{%
    \includegraphics[width=0.5\linewidth]{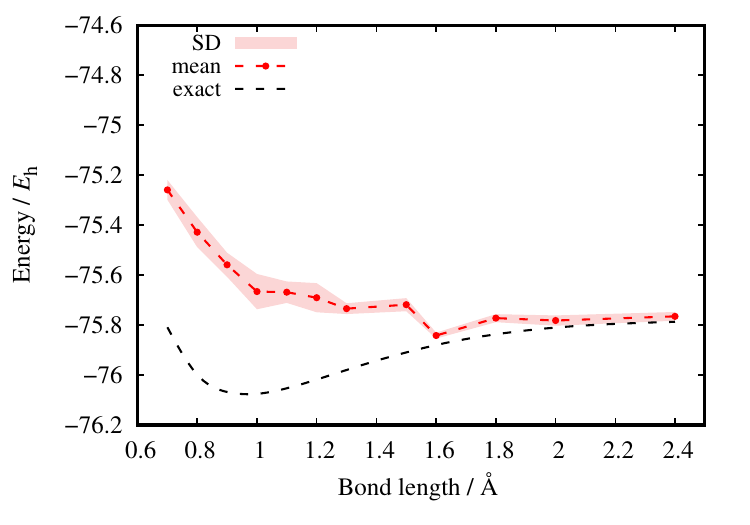}
    }    
    \caption{Potential energy curves for the symmetric stretching of H$_2$O obtained
from finite-shot evaluation of the optimized VQE state on the \texttt{ibm\_pittsburgh} QPU using 1,000 shots. Red points
and shaded regions represent the mean and $\pm 1$ standard deviation
(SD), respectively, over 10 independent runs. The dashed black line
denotes the CASSCF(4,4) reference. (a) QPU-evaluated energies;
(b) corresponding ERPA-corrected energies.}
    \label{h2o_qpu_pittsburgh}
\end{figure*}

The results obtained on the \texttt{ibm\_pittsburgh} QPU are presented in
Fig.~\ref{h2o_qpu_pittsburgh}. The raw QPU energies obtained with
1,000 measurement shots, shown in Fig.~\ref{h2o_shots_1000_qpu},
exhibit large deviations from the CASSCF reference. Application of
the ERPA correction leads to a substantial improvement, as shown in
Fig.~\ref{h2o_shots_1000_qpu_corr}. In particular, ERPA considerably
reduces both the error in the mean energies and their statistical
uncertainty. Nevertheless, the corrected energies remain far from the
reference values, with deviations reaching approximately
$400$~m$E_h$ near the equilibrium geometry. This residual error can be
attributed primarily to inaccuracies in the input 1-RDM, which, unlike
the 2-RDM, is not corrected by the present ERPA reconstruction.

\begin{figure}[!ht]
    \centering
    \includegraphics[width=\linewidth]{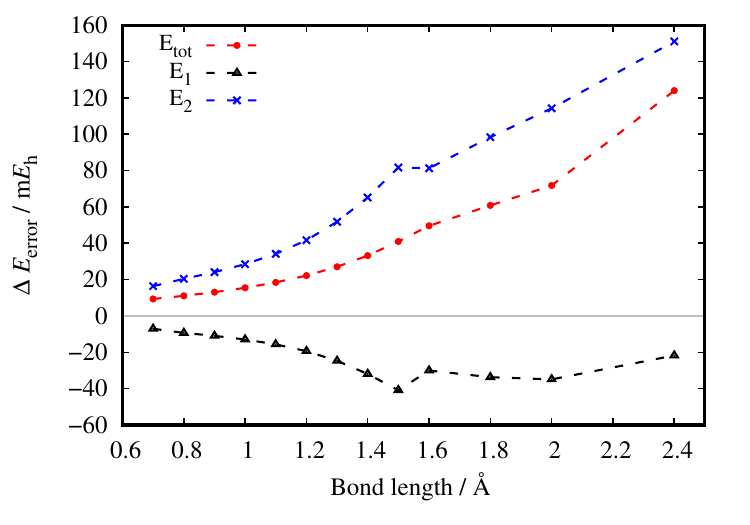}    
    \caption{Energy errors for the symmetric stretching of H$_2$O, relative to the
CASSCF(4,4) reference, obtained from a small-bond-dimension ($M=4$)
DMRG calculation. The errors are decomposed into the one-electron
($E_1$) and two-electron ($E_2$) contributions and their sum
($E_{\mathrm{tot}}$).}
    \label{approx_dmrg_wf}
\end{figure}

As pointed out above, one possibility is to replace the noisy 1-RDM
with a noise-free classical approximation. For strongly correlated
problems, such an approximation can be obtained, for example from a
low-bond-dimension DMRG calculation. One may expect one-body quantities,
such as the 1-RDM, to converge sufficiently rapidly with the bond
dimension $M$, which controls both the accuracy and computational cost
of DMRG~\cite{reiher_perspective}, even when the corresponding DMRG
energy and 2-RDM are not yet well converged. A similar idea underlies
the use of one- and two-orbital quantities obtained from inexpensive
DMRG calculations for optimizing the orbital ordering \cite{Barcza2011, Szalay2015} and guiding
active-space selection \cite{Stein2016, Golub2021}.

Figure~\ref{approx_dmrg_wf} illustrates this behavior for a
low-bond-dimension ($M=4$) DMRG calculation. Since the active space
contains only four spatial orbitals, the DMRG calculation was performed
using spin-orbital sites. The error in the one-electron contribution to
the energy is considerably smaller than that in the two-electron
contribution, the latter exceeding $150$~m$E_h$ at the most stretched
geometry. This demonstrates that the 1-RDM can already provide a
reasonably accurate description of one-body quantities even when the
DMRG wave function, and consequently the 2-RDM and total energy, remain
rather inaccurate.

\begin{figure*}[!ht]
    \centering
    \subfloat[\label{h2o_shots_1000_qpu_mps}]{%
    \includegraphics[width=0.5\linewidth]{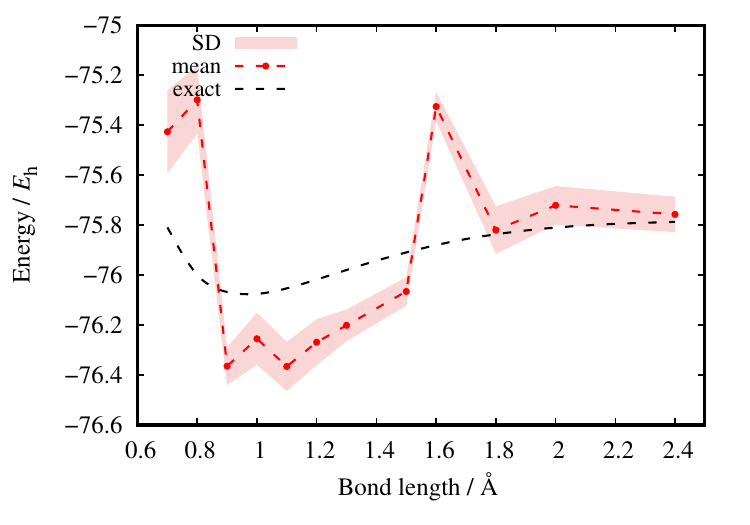}
    }
 \subfloat[\label{h2o_shots_1000_qpu_mps_corr}]{%
    \includegraphics[width=0.5\linewidth]{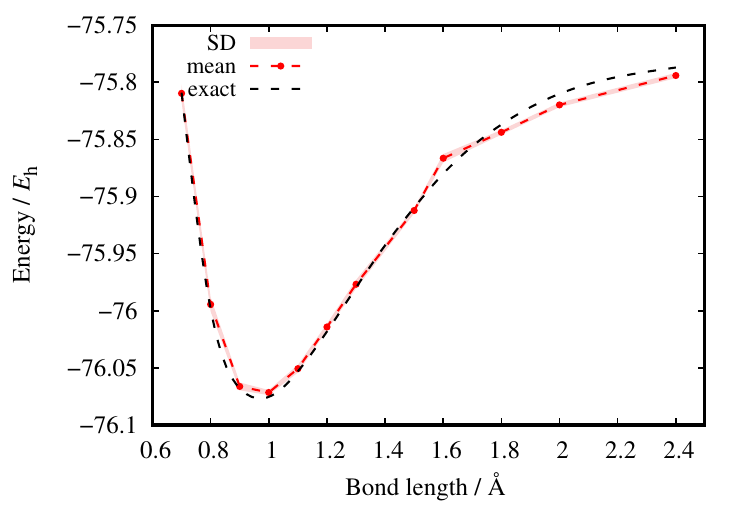}
    }    
    \caption{Potential energy curves for the symmetric stretching of H$_2$O obtained using an approximate low-$M$ DMRG 1-RDM combined with the 2-RDM evaluated on the  \texttt{ibm\_pittsburgh} QPU using 1,000 shots. Red points and shaded regions indicate the mean and $\pm 1$ standard deviation (SD), respectively, over 10 independent runs. The dashed black line denotes the CASSCF(4,4) reference. (a) Energies obtained from the combined DMRG 1-RDM and QPU-evaluated 2-RDM; (b) corresponding ERPA-corrected energies.}
    \label{h2o_qpu_pittsburgh_mps}
\end{figure*}

Importantly, simply combining this approximate DMRG 1-RDM with the
2-RDM measured on the QPU does not by itself provide accurate energies.
As shown in Fig.~\ref{h2o_shots_1000_qpu_mps}, the resulting
errors remain very large, reaching up to approximately $565$~m$E_h$ for $r =  1.6$ \AA. A
dramatically different picture emerges when the same approximate
noise-free 1-RDM is used as input to the ERPA reconstruction as shown in 
Fig.~\ref{h2o_shots_1000_qpu_mps_corr}. In this case, ERPA
effectively acts as a filter for the noisy QPU 2-RDM. Through the
contraction constraint, the reconstructed 2-RDM is forced to be
consistent with the input noise-free approximate 1-RDM. The
resulting energies are remarkably accurate, with a maximum deviation
of only about $15$~m$E_h$ from the CASSCF(4,4) reference over the
entire potential energy curve. Furthermore, comparison with the
finite-shot results in Fig.~\ref{h2o_shots_10000_corr} suggests that
the negative bias associated with the ERPA reconstruction is partially
compensated by the positive bias introduced by QPU noise, leading to
the high overall accuracy observed in
Fig.~\ref{h2o_shots_1000_qpu_mps_corr}.

\begin{figure*}[!ht]
    \centering
    \subfloat[\label{qpu_1000_ac0}]{%
    \includegraphics[width=0.5\linewidth]{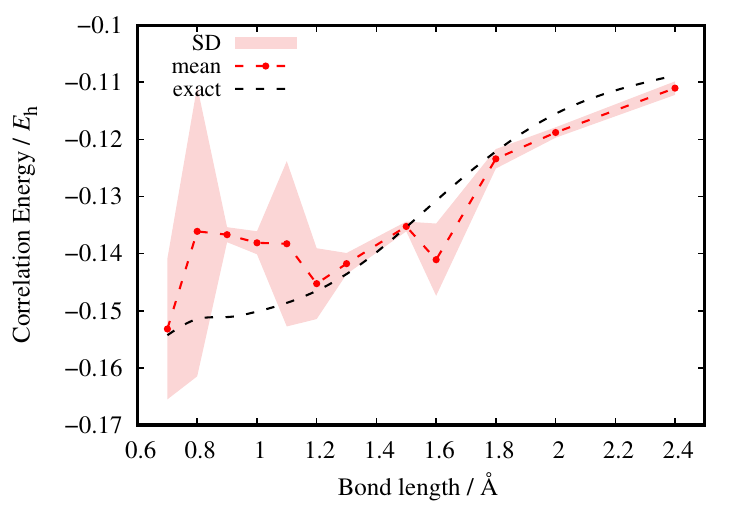}
    }
 \subfloat[\label{qpu_mps_1000_ac0}]{%
    \includegraphics[width=0.5\linewidth]{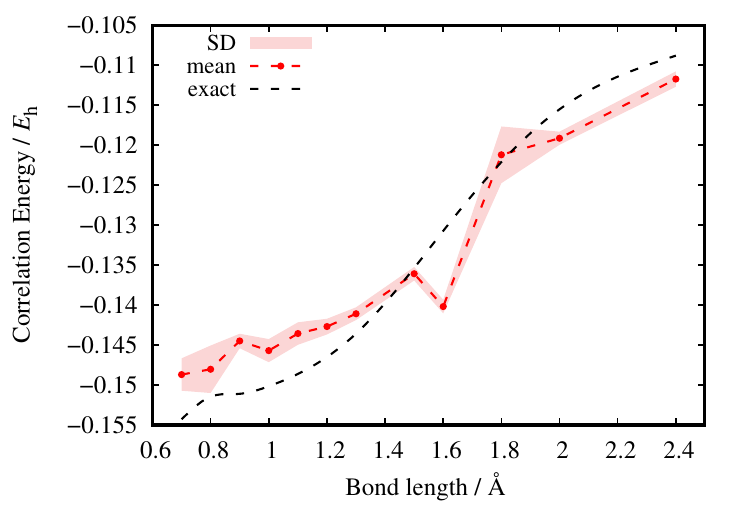}
    }    
    \caption{Out-of-CAS AC0 correlation energies for the symmetric stretching
of H$_2$O obtained using RDMs evaluated on the \texttt{ibm\_pittsburgh}
QPU with 1,000 shots. Red points and shaded regions represent the mean
and $\pm 1$ standard deviation (SD), respectively, over 10 independent
runs. The dashed black line denotes the reference AC0 correlation energy
evaluated from the CASSCF(4,4) RDMs. (a) AC0 correlation energies
obtained entirely from the QPU-evaluated RDMs; (b) AC0 correlation
energies obtained using the approximate low-$M$ DMRG 1-RDM combined
with the QPU-evaluated 2-RDM.}
    \label{h2o_ac0}
\end{figure*}

The final component of our protocol is the AC0 correction, which
accounts for the predominantly dynamical correlation missing from the
CAS description. Figure~\ref{h2o_ac0} compares AC0 correlation energies
obtained directly from the RDMs measured on the QPU
[Fig.~\ref{qpu_1000_ac0}] with those obtained using the approximate
1-RDM from the low-$M$ DMRG calculation together with the QPU-measured
2-RDM [Fig.~\ref{qpu_mps_1000_ac0}]. The latter combination yields
correlation energies in closer agreement with the reference AC0
correction evaluated from the CASSCF(4,4) RDMs. Nevertheless, even the
AC0 corrections evaluated entirely from the QPU-measured RDMs exhibit
comparatively moderate errors, particularly considering the
substantially larger errors in the underlying CAS energies. This
relative robustness can be related to the structure of the AC approach,
in which the correlation correction contains contributions associated
with both the one- and two-electron parts of the
energy~\cite{Pernal2017}. Consequently, errors in the input 1-RDM do
not propagate into the AC0 correction in the same direct manner as
they do in the ERPA-based reconstruction of the CAS energy, where the
input 1-RDM remains unchanged.

\begin{figure*}[!ht]
    \centering
    \subfloat[\label{qpu}]{%
    \includegraphics[width=0.5\linewidth]{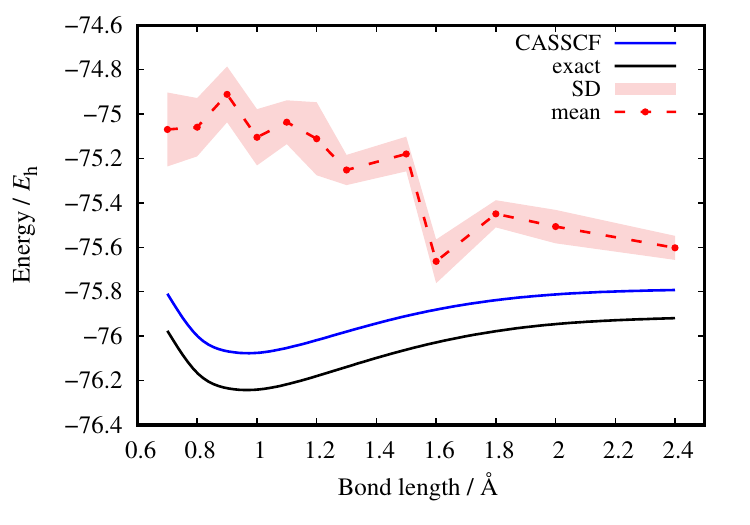}
    }
 \subfloat[\label{ac0}]{%
    \includegraphics[width=0.5\linewidth]{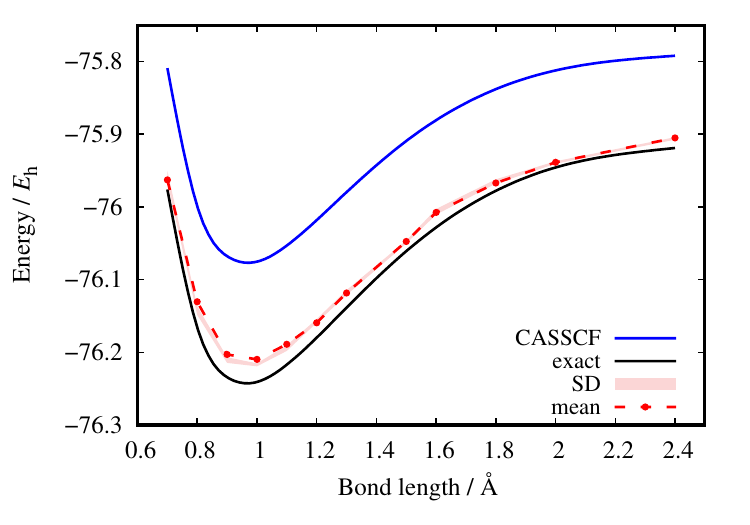}
    }    
    \caption{Comparison of total energies for the symmetric stretching of
H$_2$O. Red points and shaded regions represent the mean and $\pm 1$
standard deviation (SD), respectively, over 10 independent QPU runs.
The blue line denotes the CASSCF(4,4) energy, while the black line
denotes the essentially exact cc-pVDZ reference obtained from DMRG
calculations using DBSS with TRE of $10^{-7}$. (a) Original unmitigated QPU energies
obtained with 1,000 shots; (b) final energies obtained by combining the
ERPA-mitigated CAS energies with the AC0 correlation corrections, using
the approximate low-$M$ DMRG 1-RDM together with the QPU-evaluated
2-RDM.}
    \label{h2o_tot}
\end{figure*}

Figure~\ref{ac0} presents the final total energies obtained by combining
the ERPA-mitigated CAS energies with the corresponding AC0 correlation
corrections. The AC0 correction recovers a substantial fraction of the
dynamical correlation missing from the CAS description, bringing the
resulting energies into close agreement with the essentially exact
cc-pVDZ reference obtained from DMRG calculations employing dynamic
block-state selection (DBSS) \cite{Legeza2003} and a prescribed truncation error (TRE) of
$10^{-7}$. Overall, the combined ERPA+AC0 protocol reduces the errors
by orders of magnitude compared with the original, unmitigated QPU
results, which are shown again in Fig.~\ref{qpu} together with the
CASSCF(4,4) and exact cc-pVDZ reference energies.

\subsection{Automerization of cyclobutadiene}

As a second benchmark, we consider the automerization of cyclobutadiene, a prototypical multireference problem that poses a challenge for single-reference methods. The reaction involves the interconversion of two equivalent rectangular ($D_{2h}$) structures through a square ($D_{4h}$) transition state, as illustrated in Fig.~\ref{cb_reaction}. While the rectangular ground state has predominantly closed-shell character, the square transition state exhibits strong multireference character due to the degeneracy of the frontier $\pi$ orbitals, making the accurate description of the reaction barrier particularly challenging.

Owing to the symmetry of the reaction, only one half of the pathway was considered, connecting the $D_{2h}$ minimum to the $D_{4h}$ transition state.

Having analyzed the individual components of the error-mitigation protocol for the water molecule, we do not repeat the detailed analysis here. Instead, we focus on comparing the raw QPU results with those obtained using the combined DMRG 1-RDM, ERPA-based reconstruction of the QPU 2-RDM, and the subsequent AC0 correlation correction. Before presenting this comparison, we briefly assess the accuracy of the approximate spin-orbital DMRG ($M=4$) calculation used to provide the noise-free 1-RDM.

\begin{figure}[!ht]
    \centering
    \includegraphics[width=\linewidth]{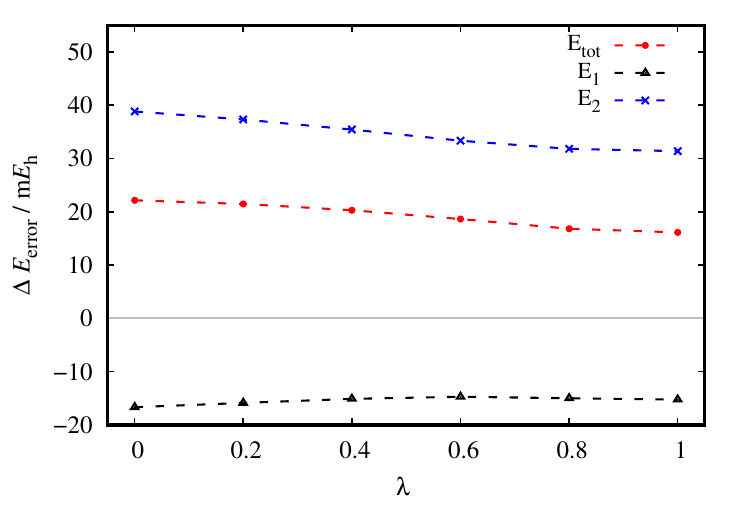}    
    \caption{Energy errors along the automerization pathway of cyclobutadiene, relative to the CASSCF(4,4) reference, obtained from a small-bond-dimension ($M=4$) DMRG calculation. The errors are decomposed into the one-electron ($E_1$) and two-electron ($E_2$) contributions and their sum ($E_{\mathrm{tot}}$). The reaction coordinate $\lambda$ ranges from 0 (rectangular $D_{2h}$ minimum) to 1 (square $D_{4h}$ transition state).}
    \label{approx_dmrg_cb}
\end{figure}

As shown in Fig.~\ref{approx_dmrg_cb}, the total DMRG energy deviates from the CASSCF(4,4) reference by approximately 16--22 m$E_h$. Interestingly, the error in the one-electron energy contribution remains nearly constant along the automerization pathway, varying by only about 2 m$E_h$, whereas the two-electron contribution exhibits a somewhat larger variation of approximately 8 m$E_h$. Thus, despite the limited accuracy of the underlying DMRG wave function, its 1-RDM provides a relatively balanced description along the reaction coordinate.

\begin{figure*}[!ht]
    \centering
    \subfloat[\label{cbd_qpu}]{%
    \includegraphics[width=0.5\linewidth]{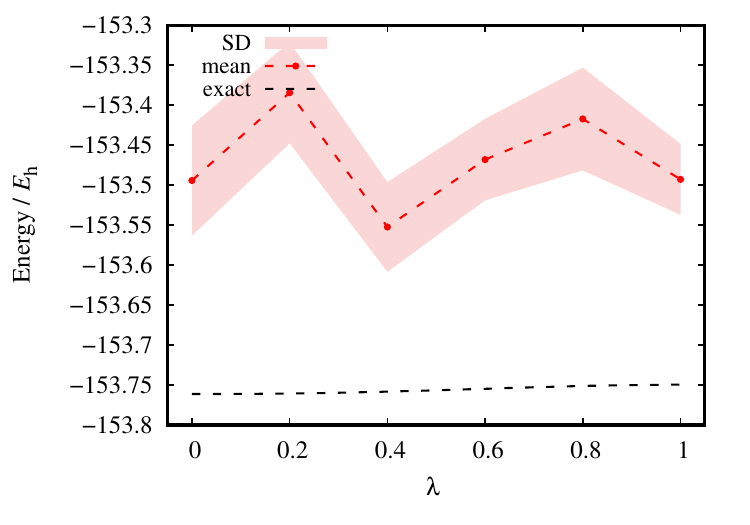}
    }
 \subfloat[\label{cbd_corr}]{%
    \includegraphics[width=0.5\linewidth]{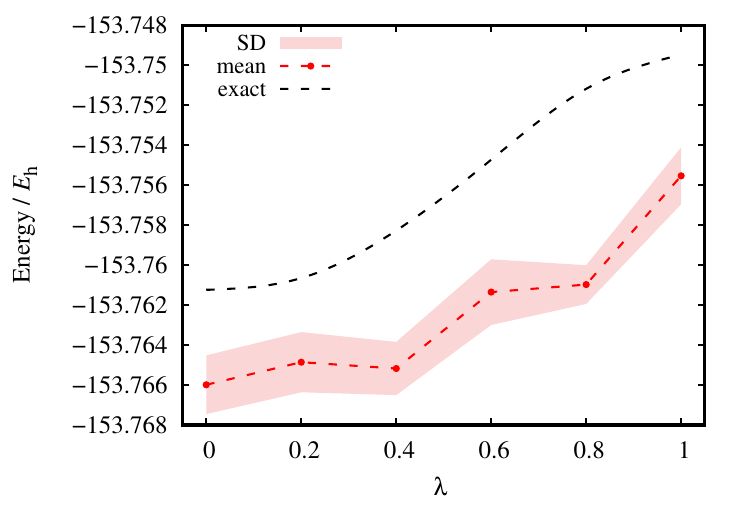}
    }    
    \caption{Potential energy curves along the automerization pathway of cyclobutadiene obtained from measurements on the \texttt{ibm\_pittsburgh} QPU using 1,000 shots. Red points and shaded regions represent the mean and $\pm 1$ standard deviation (SD), respectively, over 10 independent runs. The dashed black line denotes the CASSCF(4,4) reference. (a) Unmitigated QPU energies; (b) ERPA-mitigated energies obtained using the approximate low-$M$ DMRG 1-RDM and the ERPA-reconstructed QPU 2-RDM. The reaction coordinate $\lambda$ connects the rectangular ($D_{2h}$) minimum ($\lambda=0$) with the square ($D_{4h}$) transition state ($\lambda=1$).}
    \label{cbd}
\end{figure*}

Figure~\ref{cbd} compares the energies obtained directly from the QPU-measured RDMs [Fig.~\ref{cbd_qpu}] with those calculated using the low-$M$ DMRG 1-RDM and the ERPA-reconstructed 2-RDM, as described above [Fig.~\ref{cbd_corr}]. As for water, the improvement is dramatic, with errors reduced by orders of magnitude. Despite slight deviations from a smooth energy profile, the mitigated energies remain within 10 m$E_h$ of the CASSCF(4,4) reference throughout the automerization pathway. The corrected energies lie systematically below the reference, suggesting that the negative bias associated with the approximate 1-RDM and ERPA reconstruction is only partially compensated by the positive bias introduced by QPU noise, consistent with the behavior observed for water.

Since an essentially exact potential energy curve is not available for cyclobutadiene, we assess the performance of the complete ERPA+AC0 protocol by evaluating the automerization barrier. As shown in Table~\ref{cbd_barrier}, despite the systematic underestimation of the absolute CAS energies, the corresponding barrier differs from the CASSCF(4,4) reference by less than 1 kcal/mol. Upon including the AC0 correction, evaluated using the approximate DMRG 1-RDM and QPU-measured 2-RDM, the deviation from the CASSCF(4,4)/AC0 barrier remains below 2 kcal/mol. For comparison, the table also includes results from single-reference CCSD and multireference coupled clusters, MR AQCC \cite{EckertMaksi2006}, tailored CCSD (TCCSD)\cite{Kinoshita2005, Veis2016}, and its Hilbert space multireference generalization (MR TCCSD \cite{Demel2023}) in particular. While conventional CCSD substantially overestimates the barrier, the mitigated QPU results are in good agreement with the multireference benchmarks, demonstrating the ability of the proposed protocol to recover chemically meaningful energy differences from highly noisy quantum-hardware measurements. Experimental estimates of the barrier span a broad range of approximately 1.6--12 kcal/mol \cite{Whitman1982}, preventing a more definitive assessment against experiment.

\begin{table}[!ht]
  \centering
  \begin{tabular}{r r}
  \hline
  \hline
    & \ $\Delta E$ / kcal $\cdot$ mol$^{-1}$ \\
  \hline
  $E(\gamma^\text{MPS}, \Gamma^{\text{ERPA}})$ & 6.56 \\
  CASSCF(4,4) & 7.38 \\
  \hline
  $E(\gamma^\text{MPS}, \Gamma^{\text{ERPA}})$/AC0 & 8.99 \\
  CASSCF(4,4)/AC0 & 10.91 \\
  \hline
  CCSD\cite{Demel2023} & 22.46 \\ 
  MR AQCC\cite{EckertMaksi2006} & 8.4 \\
  TCCSD(20,19)\cite{Demel2023} & 10.99 \\
  MR TCCSD(20,19)\cite{Demel2023} & 9.13 \\
  \hline
  \hline
  \end{tabular}
  \caption{Automerization barriers of cyclobutadiene (in kcal/mol) calculated using the ERPA-mitigated QPU energies, with and without the AC0 correlation correction, compared with CASSCF(4,4), CASSCF(4,4)/AC0, and selected single- and multireference coupled-cluster methods. The QPU-based values are averaged over 10 independent runs. All calculations employ the cc-pVTZ basis set.}
  \label{cbd_barrier}
\end{table}

\section{Conclusions}

In this work, we have introduced an error-mitigation strategy for VQE calculations based on the reconstruction of the two-particle reduced density matrix (2-RDM) using the extended random phase approximation (ERPA). Since the ERPA reconstruction preserves the input one-particle reduced density matrix (1-RDM), an accurate estimate of the latter is essential. We have demonstrated that a noise-free approximate 1-RDM obtained from a low-bond-dimension DMRG calculation, combined with the ERPA reconstruction of the noisy QPU-measured 2-RDM, can reduce errors in the active-space energies by orders of magnitude. This substantial improvement was demonstrated for two strongly correlated benchmark problems, namely the symmetric stretching of water and the automerization of cyclobutadiene, using measurements performed on an IBM quantum processor.

Furthermore, the mitigated energies can be combined with the multireference adiabatic connection (AC0) approach to recover dynamical electron correlation outside the active space. The resulting potential energy curve for water closely approaches the essentially exact reference, while the automerization barrier of cyclobutadiene agrees well with established multireference calculations. Importantly, the proposed mitigation scheme requires no additional quantum resources beyond the measurement of the 1- and 2-RDMs already needed for the energy evaluation and AC0 correction. It relies entirely on classical post-processing, provided that the molecular orbitals are kept fixed, and can be combined with other error-mitigation techniques.

Overall, our results demonstrate that physically motivated reconstruction of reduced density matrices provides an effective route to improving the accuracy of quantum-hardware calculations. While the present work focuses on mitigating errors in RDMs measured from preoptimized VQE states, incorporating the proposed strategy directly into the VQE optimization procedure represents a natural direction for future research.

\section*{Acknowledgment}
This work was supported by the Czech Science Foundation (grant no.\ 25-18486S), the Grant Agency of the Czech Technical University in Prague, grant No. SGS25/165/OHK4/3T/14, the Ministry of Education, Youth and Sports of the Czech Republic through the e-INFRA CZ (ID:90254), and the National Science Center of Poland (Grant No. 2021/43/I/ST4/02250). We gratefully acknowledge the Czech Quantum Innovation Center (QIC) for providing access to IBM quantum processing units (QPUs).

\bibliography{references}

\end{document}